\newif\ifpreprint

\preprintfalse 

\ifpreprint
\documentclass[aip,jcp,amsmath,amssymb,preprint]{revtex4-1}
\else
\documentclass[aip,jcp,amsmath,amssymb,reprint]{revtex4-1}
\fi

\usepackage{xspace}
\usepackage{graphicx}
\usepackage{dcolumn}
\usepackage{bm}
\usepackage{amsfonts}
\usepackage{pbox}
\usepackage{braket}
\usepackage{multirow}
\usepackage{threeparttable}
\usepackage{float}
\usepackage{xspace}
\usepackage{color}
\usepackage{xcolor}
\usepackage{fancyvrb}
\usepackage{adjustbox}
\usepackage{standalone}
\usepackage{marvosym}
\usepackage[utf8]{inputenc}
\usepackage[note-name=,use-sort-key=false]{notes2bib}

\usepackage{hyperref}
\hypersetup{
    colorlinks,
    linkcolor={red!50!black},
    citecolor={red!70!black},
    urlcolor={red!80!black}
}

\makeatletter
\renewcommand*\env@matrix[1][*\c@MaxMatrixCols c]{%
  \hskip -\arraycolsep
  \let\@ifnextchar\new@ifnextchar
  \array{#1}}
\makeatother

\usepackage{listings}
\definecolor{codegreen}{rgb}{0.58,0.4,0.2}
\definecolor{codegray}{rgb}{0.5,0.5,0.5}
\definecolor{codepurple}{rgb}{0.25,0.35,0.55}
\definecolor{codeblue}{rgb}{0.30,0.60,0.8}
\definecolor{backcolour}{rgb}{0.98,0.98,0.98}
\definecolor{mygray}{rgb}{0.5,0.5,0.5}

\definecolor{sqred}{rgb}{0.85,0.1,0.1}
\definecolor{sqgreen}{rgb}{0.25,0.65,0.15}
\definecolor{sqorange}{rgb}{0.90,0.50,0.15}
\definecolor{sqblue}{rgb}{0.10,0.3,0.60}

\lstdefinestyle{mystyle}{
    backgroundcolor=\color{backcolour},
    commentstyle=\color{codegreen},
    keywordstyle=\color{codeblue},
    numberstyle=\tiny\color{codegray},
    stringstyle=\color{codepurple},
    basicstyle=\ttfamily\footnotesize,
    breakatwhitespace=false,
    breaklines=true,
    captionpos=b,
    keepspaces=true,
    numbers=left,
    numbersep=5pt,
    numberstyle=\ttfamily\tiny\color{mygray},
    showspaces=false,
    showstringspaces=false,
    showtabs=false,
    tabsize=2
}

\newcommand*{\np}[1]{{n^{+}_{#1}}}
\newcommand*{\nm}[1]{{n^{-}_{#1}}}
\newcommand*{\npvec}[0]{{\mathbf{n}^{+}}}
\newcommand*{\nmvec}[0]{{\mathbf{n}^{-}}}

\newcommand*{\cp}[1]{{c^{+}_{#1}}}
\newcommand*{\cm}[1]{{c^{-}_{#1}}}
\newcommand*{\cpvec}[0]{{\mathbf{c}^{+}}}
\newcommand*{\cmvec}[0]{{\mathbf{c}^{-}}}

\newcommand{\mref}[0]{\Psi_0}
\newcommand{\tens}[3]{{#1}_{#2}^{#3}}
\newcommand{\dfock}[1]{\epsilon_{#1}}
\newcommand{\cop}[1]{\hat{a}^\dag_{#1}}
\newcommand{\aop}[1]{\hat{a}^{\phantom \dag}_{#1}}
\newcommand{\sqop}[2]{\hat{a}_{#2}^{#1}}
\newcommand{\qop}[1]{\hat{q}_{#1}}
\newcommand{\aphystei}[2]{\bra{#1}\!\!\ket{#2}}
\newcommand{\kro}[2]{\delta_{#2}^{#1}}
\newcommand{\density}[2]{\gamma_{#2}^{#1}}
\newcommand{\hdensity}[2]{\eta_{#2}^{#1}}
\newcommand{\cumulant}[2]{\lambda_{#2}^{#1}}
\newcommand{\no}[1]{ \{ {#1} \}}

\newcommand{\copc}[0]{\cop{\mathbb{C}}}
\newcommand{\copa}[0]{\cop{\mathbb{A}}}

\newcommand{\aopc}[0]{\aop{\mathbb{C}}}
\newcommand{\aopa}[0]{\aop{\mathbb{A}}}

\newcommand{\wicked}[0]{\textsc{Wick\&d}\xspace}

\makeatletter 

\newbox\swb@xone
\newbox\swb@xtwo
\newbox\swb@xthree
\newbox\swb@xfour
\newdimen\swdimen@ne
\newdimen\swdimentw@

\newcommand{\acontraction}[5][1ex]{%
  \mathchoice
    {\acontraction@\displaystyle{#2}{#3}{#4}{#5}{#1}}%
    {\acontraction@\textstyle{#2}{#3}{#4}{#5}{#1}}%
    {\acontraction@\scriptstyle{#2}{#3}{#4}{#5}{#1}}%
    {\acontraction@\scriptscriptstyle{#2}{#3}{#4}{#5}{#1}}}%
\newcommand{\acontraction@}[6]{%
  \setbox\swb@xone=\hbox{${}#1{}#2{}$}%
  \setbox\swb@xtwo=\hbox{${}#1{}#3{}$}%
  \setbox\swb@xthree=\hbox{${}#1{}#4{}$}%
  \setbox\swb@xfour=\hbox{${}#1{}#5{}$}%
  \swdimen@ne=\wd\swb@xtwo%
  \advance\swdimen@ne by \wd\swb@xfour%
  \divide\swdimen@ne by 2%
  \advance\swdimen@ne by \wd\swb@xthree%
  \vbox{%
    \hbox to 0pt{%
      \kern \wd\swb@xone%
      \kern 0.5\wd\swb@xtwo%
      \acontraction@@{\swdimen@ne}{#6}%
      \hss}%
    \vskip 0.5ex
    \vskip\ht\swb@xtwo}}

\newcommand{\acontraction@@}[3][0.05em]{%
  \hbox{%
    \vrule width #1 height 0pt depth #3%
    \vrule width #2 height 0pt depth #1%
    \vrule width #1 height 0pt depth #3%
    \relax}}
\let\contraction\acontraction

\newcommand{\lcontraction}[5][1ex]{%
  \mathchoice
    {\lcontraction@\displaystyle{#2}{#3}{#4}{#5}{#1}}%
    {\lcontraction@\textstyle{#2}{#3}{#4}{#5}{#1}}%
    {\lcontraction@\scriptstyle{#2}{#3}{#4}{#5}{#1}}%
    {\lcontraction@\scriptscriptstyle{#2}{#3}{#4}{#5}{#1}}}%
\newcommand{\lcontraction@}[6]{%
  \setbox\swb@xone=\hbox{${}#1{}#2{}$}%
  \setbox\swb@xtwo=\hbox{${}#1{}#3{}$}%
  \setbox\swb@xthree=\hbox{${}#1{}#4{}$}%
  \setbox\swb@xfour=\hbox{${}#1{}#5{}$}%
  \swdimen@ne=\wd\swb@xtwo%
  \advance\swdimen@ne by \wd\swb@xfour%
  \divide\swdimen@ne by 2%
  \advance\swdimen@ne by \wd\swb@xthree%
  \vbox{%
    \hbox to 0pt{%
      \kern \wd\swb@xone%
      \kern 0.5\wd\swb@xtwo%
      \lcontraction@@{\swdimen@ne}{#6}%
      \hss}%
    \vskip 0.5ex
    \vskip\ht\swb@xtwo}}

\newcommand{\lcontraction@@}[3][0.05em]{%
  \hbox{%
    \vrule width #1 height 0pt depth #3%
    \vrule width #2 height 0pt depth #1%
    \relax}}

\makeatother 

\allowdisplaybreaks

\begin{document}

\title{Automatic derivation of fermionic many-body theories based on general Fermi vacua}


\author{Francesco A. Evangelista}
\email{francesco.evangelista@emory.edu}
\affiliation{Department of Chemistry and Cherry Emerson Center for Scientific Computation, Emory University, Atlanta, GA, 30322}

\date{\today}

\begin{abstract}
This paper describes \wicked, an implementation of the algebra of second-quantized operators normal ordered with respect to general correlated references and the corresponding Wick theorem [W. Kutzelnigg and D. Mukherjee, \textit{J. Chem. Phys.} \textbf{107}, 432
(1997)].
\wicked employs a compact representation of operators and a backtracking algorithm to efficiently evaluate Wick contractions.
Since \wicked can handle both fully and partially contracted terms, it can be applied to both projective and Fock-space many-body formalisms.
To demonstrate the usefulness of \wicked, we use it to evaluate the single-reference coupled cluster equations up to octuple excitations and report an automated derivation and implementation of the second-order driven similarity renormalization group multireference perturbation theory (DSRG-MRPT2).
\end{abstract}

\maketitle

\section{Introduction}
The formalism of second quantization, first developed in the context of the quantum field theory,\cite{Dirac.1927.10.1098/rspa.1927.0039} plays a fundamental role in many-body perturbation theory,\cite{Moller.1934.10.1103/physrev.46.618} coupled cluster methods,\cite{Coester.1958,Coester.1960.10.1016/0029-5582(60)90140-1,Cizek.1966.10.1063/1.1727484} and Green's functions approaches\cite{Bohm.1951.10.1103/physrev.82.625,Pines.1952.10.1103/PhysRev.85.338,Hedin.1965.10.1103/physrev.139.a796,Langhoff.1977,Schirmer.1982.10.1103/physreva.26.2395,Schirmer.1989.10.1063/1.457081,Holleboom.1990.10.1063/1.459578} for molecules and solids.
Two important tools for the manipulation and simplification of expressions involving second-quantized operators are Wick's theorem\cite{Wick.1950.10.1103/physrev.80.268} and diagrammatic methods.\cite{Shavitt.2009}
These are traditionally formulated using particle/hole quasiparticle operators defined with respect to a single determinant Fermi vacuum (reference state).
However, even though these tools can simplify the derivation of equations for many-body theories, this process is prone to human error and may be quite lengthy or even impossible to complete in a reasonable amount of time.

In the past three decades, computer-aided derivation of many-body equations has played an ever-increasing role in quantum chemistry.\cite{Hirata.2006.10.1007/s00214-005-0029-5}
Starting with the pioneering works of Janssen and Schaefer\cite{Janssen.1991.10.1007/bf01113327} and Li and Paldus,\cite{Paldus.1973,Li.1994.10.1063/1.468074}
the early 2000s saw the rapid development of automatic derivation and implementation tools based on algebraic,\cite{Harris.1999.10.1002/(SICI)1097-461X(1999)75:4/5<593::AID-QUA24>3.0.CO;2-H,Nooijen.2001.10.1016/s0166-1280(01)00475-4,Nooijen.2002.10.3390/i3060656,Hirata.2003.10.1021/jp034596z,Piecuch.2005.10.1002/qua.20753,Auer.2006.10.1080/00268970500275780,Kong.2009.10.1063/1.3089302,Zitko.2011.10.1016/j.cpc.2011.05.013} diagrammatic\cite{Kallay.2001.10.1063/1.1383290,Kallay.2002.10.1063/1.1483856,Bochevarov.2004.10.1063/1.1774977,Shiozaki.2008.10.1039/b803704n} and string or determinant-based methods.\cite{Olsen.2000.10.1063/1.1290005,Hirata.2000.10.1016/s0009-2614(00)00387-0,Kallay.2000.10.1063/1.481925,Evangelista.2006.10.1063/1.2357923,Sorensen.2016.10.1080/00268976.2016.1195926}
Recently, automatic derivation has been extended into many new directions, including arbitrary order response and derivatives,\cite{Ringholm.2014.10.1002/jcc.23533,Abbott.2021.10.1021/acs.jpclett.1c00607}
systems with coupled fermionic and bosonic degrees of freedom,\cite{Rubin.2021.10.1080/00268976.2021.1954709}
more general vacua (Hartree--Fock--Bogoliubov, Bardeen-Cooper-Schrieffer, or antisymmetrized geminal power states)\cite{Zhao.PhD.2018,Drudge}
 nuclear structure theory,\cite{Arthuis.2019.10.1016/j.cpc.2018.11.023,Arthuis.2021.10.1016/j.cpc.2020.107677,Tichai.2022.10.1140/epja/s10050-021-00621-6}  and the manipulation of quantum circuits.\cite{McClean.2020.10.1088/2058-9565/ab8ebc}
Several works have also investigated the problem of  optimizing tensor operations (factorization, global optimization of the contraction order, the identification of reusable intermediates, the identification of common factors).\cite{Engels-Putzka.2011.doi:10.1063/1.3561739,Pfeifer.2013.10.1103/PhysRevE.90.033315,Kats.2013.10.1063/1.4798940,Epifanovsky.2013.10.1002/jcc.23377}

In the case of multireference (MR) many-body methods,\cite{Lyakh.2012.10.1021/cr2001417,Kohn.2013.10.1002/wcms.1120,Evangelista.2018.10.1063/1.5039496} automatic derivation may be crucial to achieving a numerical implementation, as in many cases the underlying equations are too complicated for manual derivation and implementation.
Examples of automated implementations of multireference theories include equation-of-motion multireference coupled cluster (MRCC),\cite{Kong.2009.10.1063/1.3089302} arbitrary-order Mukherjee's state-specific MRCC,\cite{Das.2010.10.1063/1.3515478} internally-contracted MRCC,\cite{Evangelista.2011.10.1063/1.3559149,Hanauer.2011.doi:10.1063/1.3592786,Hanauer.2012.10.1063/1.4757728,Kohn.2020.10.1080/00268976.2020.1743889,Krupicka.2017.10.1002/jcc.24833} and analytic energy gradients and derivative couplings of multireference perturbation theories.\cite{MacLeod.2015.10.1063/1.4907717,Park.2017.10.1021/acs.jctc.7b00018}

An important tool for the derivation of equations of internally-contracted multireference theories\cite{Mukherjee.1977.10.1080/00268977700100871,Lindgren.1978.10.1002/qua.560140804,Lindgren.1985.10.1088/0031-8949/32/4/009,Banerjee.1981,Yanai.2006.10.1063/1.2196410,Yanai.2007.10.1063/1.2761870,Neuscamman.2009.10.1063/1.3086932} is the generalized normal ordering approach by Mukherjee and Kutzelnigg (MK).\cite{Mukherjee.1997.10.1016/s0009-2614(97)00714-8,Kutzelnigg.1997.10.1063/1.474405}
Mukherjee and Kutzelnigg proposed an extension of the definition of normal-ordered operators and Wick's theorem that applies to general correlated Fermi vacua.
The MK formalism provides the theoretical framework for extending Fock space or ``many-body equations'' methods\cite{Kutzelnigg.1982,Kutzelnigg.1983,Kutzelnigg.1984,Nooijen.1996.10.1063/1.470988} to the multireference case.
Later works have formalized proofs of the MK formalism\cite{Kong.2010.10.1063/1.3439395,Misiewicz.2020.10.1021/acs.jctc.0c00422} and examined the issues of spin adaptation.\cite{Shamasundar.2009.10.1063/1.3256237,Kutzelnigg.2010.10.1080/00268970903547926}
The MK approach is foundational to several methods, including canonical transformation theory,\cite{Yanai.2007.10.1063/1.2761870}  explicitly correlated basis set incompleteness corrections,\cite{Torheyden.2009.10.1063/1.3254836,Kong.2011.doi:10.1063/1.3664729}  various equation-of-motion multireference coupled cluster methods,\cite{Kong.2009.10.1063/1.3089302,Datta.2011.doi:10.1063/1.3592494,Datta.2012.10.1063/1.4766361,Nooijen.2014.10.1063/1.4866795,Margocsy.2021.10.1021/acs.jctc.1c00730} and similarity renormalization group approaches.\cite{Hergert.2013.10.1103/physrevlett.110.242501,Li.2016.10.1063/1.4947218,Li.2019.10.1146/annurev-physchem-042018-052416}

This paper describes an efficient approach to implement the MK version of Wick's theorem that combines a compact representation of operator contractions with a backtracking algorithm.
Our representation of operators is related and generalizes diagrammatic approaches used by others.\cite{Kallay.2001.10.1063/1.1383290,Hanauer.2011.doi:10.1063/1.3592786,Kohn.2020.10.1080/00268976.2020.1743889}
The algorithm described in this work is implemented in the Wick's theorem and diagrammatic code \wicked (which we pronounce ``wicked''), an open-source package developed by us and available from \textsc{GitHub}.\cite{wicked}
\wicked offers similar functionality to other software designed for the derivation of multireference theories, including Neuscamman's \textsc{SQA} package,\cite{Neuscamman.2009.10.1063/1.3125004,Saitow.2013.10.1063/1.4816627}
Valeev's \textsc{SeQuant2},\cite{SeQuant2}
Kong and Nooijen's automatic program generator \textsc{APG},\cite{Kong.PhD.2013}  K{\"o}hn's \textsc{GeCCo} program,\cite{Kohn.2020.10.1080/00268976.2020.1743889,GeCCo}
Shiozaki's \textsc{Smith},\cite{Shiozaki.2008.10.1039/b803704n,Smith3}
and the \textsc{ADG} program.\cite{Arthuis.2019.10.1016/j.cpc.2018.11.023,Arthuis.2021.10.1016/j.cpc.2020.107677,Tichai.2022.10.1140/epja/s10050-021-00621-6} 
However, some of the features unique to \wicked include the ability to generate both projective and Fock-space (many-body) equations and the support for an arbitrary number of orbital subspaces.

The article is organized as follows. In Sec.~\ref{sec:theory} we summarize the MK general normal ordering formalism and the corresponding Wick's theorem.
Section~\ref{sec:implementation} describes a general strategy to implement the MK general normal ordering and Wick's theorem.
Example applications of \wicked to single-reference and multireference methods are reported in Sec.~\ref{sec:examples}.
In Sec.~\ref{sec:discussion} we conclude with a discussion of the main features, limitations, and future extension of \wicked.

\section{Theory}
\label{sec:theory}

\subsection{Synopsis of the generalized normal ordering formalism}

In this section, we summarize the main results of the Mukherjee--Kutzelnigg normal ordering formalism for general vacua and the corresponding Wick's theorem.\cite{Mukherjee.1997.10.1016/s0009-2614(97)00714-8,Kutzelnigg.1997.10.1063/1.474405,Kong.2010.10.1063/1.3439395}
We follow the convention of writing products of second-quantized operators using the compact notation
\begin{equation}
\sqop{pq\cdots}{rs\cdots} = \cop{p} \cop{q} \cdots \aop{s} \aop{r},
\end{equation}
where upper (lower) indices correspond to creation (annihilation) operators and are read from left to right (right to left).

Consider an arbitrary $N$-electron reference state $\mref$.
Mukherjee and Kutzelnigg define a normal-ordered operator product $\no{\sqop{pq\cdots}{rs\cdots}}$ to satisfy the condition
\begin{equation}
\braket{\mref|\no{\sqop{pq\cdots}{rs\cdots}} |\mref} = 0.
\end{equation}
This definition can be applied recursively---starting from single substitution operators $\no{\sqop{p}{s}}$---to express normal-ordered operators in terms of bare operators and reduced density matrices of the reference state $\density{pq\cdots}{rs\cdots} = \braket{\mref|\sqop{pq\cdots}{rs\cdots} |\mref}$.\cite{Misiewicz.2020.10.1021/acs.jctc.0c00422}
A product of second-quantized operators $\qop{1} \qop{2} \cdots$, with $\qop{i} \in \{ \cop{p} \} \cup \{ \aop{p} \}$, can be expressed as a sum of normal-ordered terms using a generalization of Wick's theorem:
\begin{equation}
\label{eq:wick1}
\begin{split}
       \qop{1} \qop{2} \cdots = & \no{\qop{1} \qop{2} \cdots}
    + \sum_{\text{single} \atop \text{pairs}} \no{ 
        \contraction{}{\hat{q}}{\quad\quad}{}
        \qop{1} \qop{2} \cdots} \\
    &+ \sum_{\text{double} \atop \text{pairs}} \no{
        \contraction{}{\hat{q}}{\quad \;\; q}{}
        \contraction[0.75em]{\qop{1}}{\hat{q}}{\quad q}{}
        \qop{1} \qop{2} \cdots}
    + \sum_{\text{single} \atop \text{4-leg}} \no{
        \contraction{}{\hat{q}}{\quad q}{}
        \contraction{\qop{1}}{\hat{q}}{\quad q}{}
       \qop{1} \qop{2} \cdots} \\
        &
        + \sum_{\text{triple} \atop \text{pairs}}  \no{
        \contraction{}{\hat{q}}{\quad \;\; q}{}
        \contraction[0.75em]{\qop{1}}{\hat{q}}{\quad q}{}
        \contraction[1.05em]{\qop{1}\qop{2}\!}{\hat{q}}{\quad \,}{}        
        \qop{1} \qop{2} \cdots \;}
%
        + \sum_{\text{single} \atop \text{pairs}}  \sum_{\text{single} \atop \text{4-leg}} \no{
        \contraction{}{\hat{q}}{\quad q q}{}
        \contraction{\qop{1}}{\hat{q}}{\quad  q q}{}
        \contraction[0.75em]{\qop{1}\qop{2}}{\hat{q}}{q}{}
        \qop{1} \qop{2} \; \cdots \;}
    + \ldots .
\end{split}
\end{equation}
When compared to Wick's theorem for a Slater determinant reference (e.g., see Refs.~\citenum{Crawford.2000.10.1002/9780470125915.ch2} and \citenum{Shavitt.2009}) the generalized form contains two new aspects.
Firstly, pairwise contractions yield elements of the one-particle ($\boldsymbol{\gamma}_1$) or one-hole ($\boldsymbol{\eta}_1$) density matrices:
\begin{align}
    \contraction{}{\hat{a}}{{}^{p}}{\hat{a}}
    \cop{p} \aop{q}
    &= \density{p}{q} \equiv \braket{\mref| \sqop{p}{q}| \mref},\\
    \contraction{}{\hat{a}}{{}_{q}}{\hat{a}}
    \aop{q} \cop{p}
    &= \hdensity{p}{q} = \kro{p}{q} - \density{p}{q}.
\end{align}
Secondly, in addition to pairwise contractions, new multi-legged contractions appear.
A $2k$-leg contraction ($k\geq2$) involves $k$ creation and $k$ annihilation operators and corresponds to elements of the $k$-body density cumulant ($\boldsymbol{\lambda}_k$) of the reference $\mref$.\cite{Mukherjee.1997.10.1016/s0009-2614(97)00714-8,Mazziotti.1998.10.1016/s0009-2614(98)00470-9,Mukherjee.2001.10.1063/1.1337058}
For example, the following 4-leg contraction evaluates to an element of the 2-body density cumulant $\cumulant{pq}{rs}$:
\begin{equation}
\label{eq:2cum}
\begin{split}
\contraction{}{\hat{a}}{_p \cop{q} \aop{s}}{\hat{a}}
\contraction{\cop{p}}{\hat{a}}{_q}{\hat{a}}
\cop{p} \cop{q} \aop{s} \aop{r}
= \cumulant{pq}{rs} = \density{pq}{rs} - \density{p}{r} \density{q}{s} + \density{p}{s} \density{q}{r}.
\end{split}
\end{equation}
Multi-leg contractions are antisymmetric with respect to permutations of the operators involved, for example:
\begin{equation}
\contraction{}{\hat{a}}{_p \cop{q} \aop{s}}{\hat{a}}
\contraction{\cop{p}}{\hat{a}}{_q}{\hat{a}}
\cop{p} \cop{q} \aop{s} \aop{r}
=
-\contraction{}{\hat{a}}{_p \cop{s} \aop{q}}{\hat{a}}
\contraction{\cop{p}}{\hat{a}}{_s}{\hat{a}}
\cop{p} \aop{s} \cop{q}  \aop{r}
= \ldots .
\end{equation}

Two important simplifications apply to complete active space (CAS) references.
Firstly, since cumulants are zero when one or more indices correspond to empty or fully occupied spinorbitals, multi-leg contractions only connect operators labeled by active indices.
Secondly, the one-particle density matrix is block diagonal, with the occupied (unoccupied) block equal to the identity (zero) matrix.

A second Wick theorem helps express a product of normal-ordered operators  $\no{\hat{A}} \no{\hat{B}} \cdots \no{\hat{Z}}$ as a single normal-ordered product $\no{\hat{A}\hat{B}\cdots \hat{Z}}$ plus a sum of contractions:
\begin{equation}
\label{eq:wick2}
\begin{split}
    \no{\hat{A}} \no{\hat{B}} \cdots \no{\hat{Z}} =& \no{\hat{A}\hat{B}\cdots\hat{Z}}
    + \sum_{\text{single} \atop \text{pairs}} \no{ 
        \contraction{}{\hat{A}}{\,\,}{\hat{B}}
        \hat{A}\,\,\hat{B} \cdots \hat{Z}} \\
    &+ \sum_{\text{double} \atop \text{pairs}} \no{
        \contraction{\!}{\hat{A}}{\,\,}{\hat{B}}
        \contraction[0.75em]{\,}{\hat{A}}{\,\,}{\hat{B}}
        \hat{A}\,\,\hat{B}\cdots \hat{Z}}
    + \sum_{\text{single} \atop \text{4-leg}} \no{
        \contraction{\!}{\hat{A}}{\,\,}{\hat{B}}
        \contraction{\,}{\hat{A}}{\quad\cdots}{\hat{B}}
        \hat{A}\,\,\hat{B}\cdots \hat{Z}} \\
        &+ \sum_{\text{triple} \atop \text{pairs}} \no{
        \contraction{\!}{\hat{A}}{\,\,}{\hat{B}}
        \contraction[0.75em]{\,}{\hat{A}}{\,\,\quad}{\hat{B}}
        \contraction[1.05em]{\hat{A}\;\;}{\,\, \hat{B}}{\quad}{\hat{Z}}
        \hat{A}\,\,\hat{B}\cdots \hat{Z}} + \ldots .
\end{split}
\end{equation}
Since the starting operators are normal ordered, Eq.~\eqref{eq:wick2} excludes contractions that exclusively involve second-quantized operators within a normal-ordered group.
This second form of Wick's theorem plays a central role in the derivation of expressions of internally-contracted multireference theories.

\section{Implementation of Wick's theorem}
\label{sec:implementation}

In this section, we describe a general procedure to evaluate Eq.~\eqref{eq:wick2} when the orbital space is partitioned into an arbitrary number of subspaces.
We begin by defining three types of orbital subspaces and then proceed to describe a canonical representation of operators, a directed hypergraph (diagrammatic) representation of Wick contractions, and a backtracking approach for generating all unique Wick contractions.

\subsection{Orbital subspaces and reference types}

To describe the structure of the general reference state $\mref$, we partition the set of orthonormal spinorbitals $\mathbb{S} = \{ \psi_1, \psi_2, \ldots \}$ into $s$ disjoint sets $\mathbb{S}_k$ (orbital subspaces), such that
\begin{equation}
\mathbb{S} = \cup_{k=1}^{s} \mathbb{S}_k.
\end{equation}
The structure of the reference $\mref$ is defined by constraints on the occupation of the orbitals in each subspace.
We consider three types of orbital subspaces:
\begin{enumerate}
\item \textbf{Occupied}. All spinorbitals in this subspace are occupied by one electron.
\item \textbf{General}. These spinorbitals are partially occupied in the reference, and consequently, the corresponding density matrices and cumulants are nontrivial.
\item \textbf{Unoccupied}. All spinorbitals in this subspace are empty.
\end{enumerate}
The restrictions on the one-body density matrix, hole density matrix, and cumulants for each of these subspaces are reported in Tab.~\ref{tab:space_defs}.
In \wicked, the reference state is specified by the number of orbital subspaces and their type.
This information is provided by the user and it is fully customizable.
For example, a mean-field reference wave function (a single Slater determinant), is specified by partitioning $\mathbb{S}$ into occupied ($\mathbb{O}$, occupied) and virtual ($\mathbb{V}$, unoccupied) orbitals.
A CAS reference is instead specified by splitting $\mathbb{S}$ into core ($\mathbb{C}$, occupied), active ($\mathbb{A}$, general), and virtual ($\mathbb{V}$, unoccupied) sets.

\begingroup
\setlength{\tabcolsep}{10pt}
\begin{table}[htbp]
   \centering
   \caption{Definition of the orbital subspaces handled by \wicked.}   
   \begin{tabular}{@{} lccc @{}}
   \hline 
   
   \hline
      Subspace & $\density{p}{q}$ & $\hdensity{p}{q}$ & $\cumulant{pq\cdots}{rs\cdots}$ \\
      \hline
      Occupied & $\kro{p}{q}$ & 0 & 0 \\
      General &  $\density{p}{q}$ & $\hdensity{p}{q}$ & $\cumulant{pq\cdots}{rs\cdots}$ \\ 
      Unoccupied & 0 & $\kro{p}{q}$ & 0 \\            
   \hline
   
   \hline      
   \end{tabular}
   \label{tab:space_defs}
\end{table}
\endgroup

\subsection{Canonical form of operators and their representation}

The automatic enumeration of Wick contractions benefits from expressing all normal-ordered operators in a canonical form.
Then, an operator can be identified uniquely by the number of second-quantized operators that create and annihilate particles in each orbital subspace.
To express this canonical form in a compact way, we first introduce a convenient notation for products of second-quantized operators. 
We write a product of $\np{k}$ second-quantized creation operators that act on subspace $\mathbb{S}_k$ as
\begin{equation}
\underbrace{\cop{p_1} \cop{p_2} \cdots }_{\np{k}} = \sqop{\mathcal{P}_k}{},
\end{equation}
where $\mathcal{P}_k = (p_1, \ldots, p_{\np{k}})$ is a multi-index.
Similarly, we define a product of $\nm{k}$ annihilation operators in subspace $\mathbb{S}_k$ as
\begin{equation}
\underbrace{\cdots \aop{q_2} \aop{q_1}}_{\nm{k}} = \sqop{}{\mathcal{Q}_k},
\end{equation}
where $\mathcal{Q}_k$ is the multi-index $\mathcal{Q}_k = (q_1, \ldots, q_{\nm{k}})$.
This notation allows us to define a product of second-quantized operators in canonical order as
\begin{equation}
\label{eq:notation}
\underbrace{\mathcal{P}_1}_{\np{1}}
\underbrace{\mathcal{P}_2}_{\np{2}}
\cdots
\underbrace{\mathcal{Q}_2}_{\nm{2}}
\underbrace{\mathcal{Q}_1}_{\nm{1}}
\equiv
\sqop{\mathcal{P}_1\mathcal{P}_2\cdots}{\mathcal{Q}_1\mathcal{Q}_2\cdots},
\end{equation}
where groups of creation (annihilation) operators are ordered according to increasing (decreasing) subspace index.

Using the notation of Eq.~\eqref{eq:notation}, we write a normal-ordered operator $\hat{\Omega}$ in canonical form as
\begin{equation}
\label{eq:canonical_operator}
\hat{\Omega} = \frac{1}{\np{1}! \np{2}! \cdots \nm{1}! \nm{2}! \cdots}
\sum_{\mathcal{P}_1\mathcal{P}_2\cdots}  \sum_{\mathcal{Q}_1\mathcal{Q}_2\cdots}
\tens{\omega}{\mathcal{P}_1\mathcal{P}_2\cdots}{\mathcal{Q}_1\mathcal{Q}_2\cdots}
\no{\sqop{\mathcal{P}_1\mathcal{P}_2\cdots}{\mathcal{Q}_1\mathcal{Q}_2\cdots}},
\end{equation}
where $\tens{\omega}{\mathcal{P}_1\mathcal{P}_2\cdots}{\mathcal{Q}_1\mathcal{Q}_2\cdots}$ is a tensor antisymmetric with respect to separate permutations of upper and lower indices, while the numerical prefactor accounts for equivalent terms.
The term in Eq.~\eqref{eq:canonical_operator} corresponds to a vertex labeled by a label (``$\omega$'') and an operator matrix $\mathbf{N} = [\npvec \; \nmvec]$,  where $\npvec = [\np{1},\ldots,\np{s}]$ and $\nmvec = [\nm{1},\ldots,\nm{s}]$ are column vectors that define the number of creation and annihilation operators in each orbital subspace, respectively.
In this article, we represent such an object with a matrix and a label:
\begin{equation}
\hat{\Omega}
\leftrightarrow
\begin{array}{c}
\mathbf{N} \\
\omega
\end{array}
=
\begin{array}{c}
\begin{bmatrix}
\np{1} & \nm{1} \\
\vdots & \vdots \\
\np{s} & \nm{s}
\end{bmatrix} \\
\omega
\end{array}
\begin{array}{c}
\!\!
\begin{matrix}
\leftarrow \mathbb{S}_1 \\
\vdots \\
\leftarrow \mathbb{S}_s
\end{matrix} \\
\phantom{\omega}
\end{array}
\end{equation}


Next, we provide some examples to illustrate how this canonical representation of operators works.
For a CAS reference, $\mathbb{S}$ is partitioned  into core, active, and virtual sets, $\mathbb{S} = \mathbb{C} \cup \mathbb{A} \cup \mathbb{V}$.
Then, a single excitation operator that promotes an electron from a core to an active orbital is represented by a $3 \times 2$ matrix:
\begin{equation}
\label{eq:operator_matrix_example_1}
\hat{T}_{\mathbb{AC}} = 
\sum_{m}^{\mathbb{C}} \sum_{u}^{\mathbb{A}} t^{m}_{u} \no{ {\color{sqred}\cop{u}} {\color{sqblue}\aop{m}}}
\leftrightarrow
\begin{array}{c}
\begin{bmatrix}
0 & {\color{sqblue}1} \\
{\color{sqred}1} & 0 \\
0&0
\end{bmatrix} \\
t
\end{array}
\begin{array}{c}
\!\!
\begin{matrix}
\leftarrow \mathbb{C} \\
\leftarrow \mathbb{A} \\
\leftarrow \mathbb{V}
\end{matrix} \\
\phantom{t}
\end{array}
,
\end{equation}
where the corresponding second-quantized operators and entries in the matrix representation are indicated with the same color.
Note that we labeled this operator with the orbital subspace labels ($\mathbb{AC}$) corresponding to the sequence of second-quantized operators ($\cop{u} \aop{m}$).
The two-electron operator corresponding to the replacement $\psi_v\psi_u \rightarrow \psi_m \psi_e$ (with $\psi_u,\psi_v \in \mathbb{A}$, $\psi_m \in \mathbb{C}$, and $\psi_e \in \mathbb{V}$) is represented by the matrix:
\begin{equation}
\label{eq:operator_matrix_example_2}
\hat{V}_{\mathbb{CVAA}}
=
\frac{1}{2}
\sum_{m}^{\mathbb{C}}
\sum_{uv}^{\mathbb{A}}
\sum_{e}^{\mathbb{V}}
\tens{v}{me}{uv}
\no{
{\color{sqorange}\cop{m}}
{\color{sqblue}\cop{e}}
{\color{sqred}\aop{v} \aop{u}} }
\leftrightarrow
\begin{array}{c}
\begin{bmatrix}
{\color{sqorange}1} & 0  \\
0 & {\color{sqred}2} \\
{\color{sqblue}1} & 0
\end{bmatrix} \\
v
\end{array}
,
\end{equation}
where $\tens{v}{me}{uv} = \aphystei{me}{uv}$ is an antisymmetrized two-electron integral in physicist notation.
The prefactor $1/2$ accounts for the equivalent contributions of terms in which the indices $u$ and $v$ are exchanged.

The same notation may be extended to define a product of operators.
In this case we just arrange the matrices according to the order of the operators and assign a numerical prefactor to the entire expression.
For example, the product $
\hat{V}_{\mathbb{CVAA}}
\frac{1}{2}\hat{T}_{\mathbb{AC}}^2$ is represented as an ordered list of operator matrices and labels multiplied by the scalar factor $1/2$:
\begin{equation}
\hat{V}_{\mathbb{CVAA}}
\frac{1}{2}\hat{T}_{\mathbb{AC}}^2
\leftrightarrow
\frac{1}{2}
\begin{array}{ccc}
\begin{bmatrix}
1 & 0  \\
0 & 2 \\
1 & 0
\end{bmatrix} &
\begin{bmatrix}
0 & 1 \\
1 & 0 \\
0&0
\end{bmatrix} &
\begin{bmatrix}
0 & 1 \\
1 & 0 \\
0&0
\end{bmatrix}
\\
v & t & t
\end{array}.
\end{equation}

\subsection{Representation of contractions}

\begin{figure}[h!]
   \centering
   \includegraphics{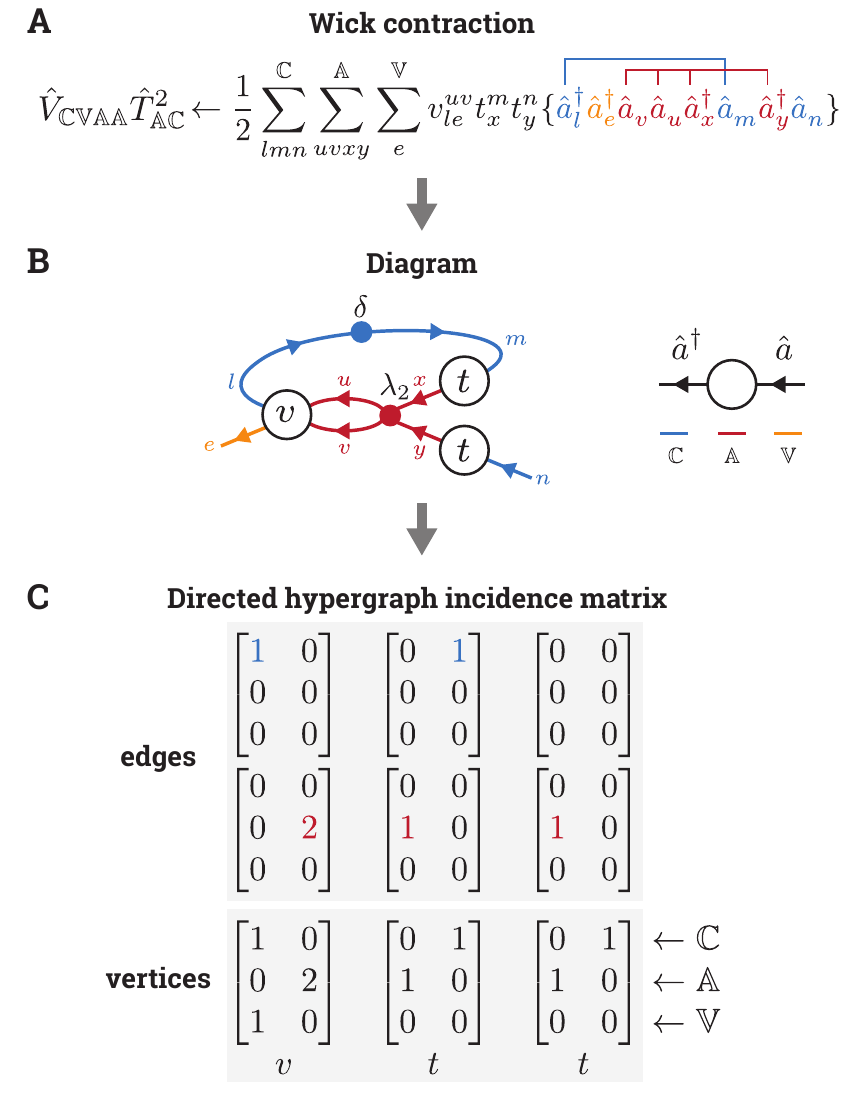}
   \caption{Illustration of how Wick contractions are represented in \wicked. Here we consider one term contributing to $\hat{V}_{\mathbb{CVAA}}\hat{T}_{\mathbb{AC}}^2$. (A) Algebraic representation of a Wick contraction. (B) Diagrammatic representation of the Wick contraction (a directed hypergraph). (C) Wick contraction encoded as an incidence matrix where the vertices and edges represent operators and elementary contractions, respectively. Arrows pointing in (out of) a vertex correspond to annihilation (creation) operators.}
   \label{fig:hypergraph}
\end{figure}

Having defined a canonical representation for normal-ordered operators, we proceed to define a canonical representation of Wick contractions [Eq.~\eqref{eq:wick2}].
As shown in Fig.~\ref{fig:hypergraph}A, a Wick contraction expressed in algebraic form may be represented as a diagram, that is, a graph in which vertices correspond to normal-ordered operators and edges represent contractions.
In this interpretation, contractions (edges) can connect an arbitrary number of operators and they encode the type of second-quantized operator connected (creation or annihilation); therefore, one can establish a correspondence between diagrams and directed hypergraphs (Fig.~\ref{fig:hypergraph}B).
One way to represent a directed hypergraph is via an incidence matrix (Fig.~\ref{fig:hypergraph}C) that encodes how the edges (contractions) connect to the vertices.

When discussing Wick contractions, we distinguish between elementary (connected) and composite (disconnected) contractions.
Elementary contractions are individual $2k$-leg contractions involving two or more second-quantized operators and will be denoted as $\mathcal{C}_i$.
To represent contractions we use a notation similar to the one used for normal-ordered operators.
For each operator, we represent an elementary contraction with a matrix $\mathbf{C} = [\cpvec \; \cmvec]$,  where $\cpvec = [\cp{1},\ldots,\cp{s}]$ and $\cmvec = [\cm{1},\ldots,\cm{s}]$ are column vectors that define the number of creation and annihilation operators contracted in each orbital subspace, respectively.
Then, an elementary contraction can be represented as a list of contraction matrices:
\begin{equation}
\mathcal{C} = 
\mathbf{C}_1 \mathbf{C}_2 \cdots
\end{equation}

For example, consider the single-reference partitioning $\mathbb{S} = \mathbb{O} \cup \mathbb{V}$ and the product 
\begin{equation}
\frac{1}{4}\sum_{ijk}^{\mathbb{O}} \sum_{abc}^{\mathbb{V}} \tens{f}{k}{c}\tens{t}{ab}{ij}
\no{\cop{k} \aop{c}} \no{\cop{a} \cop{b} \aop{j} \aop{i} }.
\end{equation}
The single contraction between the leftmost creation operator $\cop{k}$ (red) and the rightmost annihilation operator $\aop{i}$ (blue) is represented in the following way:
\begin{equation}
\frac{1}{4}\sum_{ijk}^{\mathbb{O}} \sum_{abc}^{\mathbb{V}} \tens{f}{k}{c}\tens{t}{ab}{ij}
\contraction{\{}{\hat{a}}{{}_k \aop{c} \cop{a} \cop{b} \aop{j}}{a}
\no{{\color{sqred}\cop{k}} \aop{c} \cop{a} \cop{b} \aop{j} {\color{sqblue}\aop{i}}}
\leftrightarrow
\begin{array}{cc}
\begin{bmatrix}
{\color{sqred}1} & 0 \\
0 & 0
\end{bmatrix}
&
\begin{bmatrix}
0 & {\color{sqblue}1} \\
0 & 0
\end{bmatrix} \\[15pt]
\begin{bmatrix}
1& 0 \\
0 & 1
\end{bmatrix}
&
\begin{bmatrix}
0 & 2 \\
2 & 0
\end{bmatrix} \\
f & t
\end{array}
\label{eq:example_elementary}
\end{equation}
This representation does not specify the operators contracted  within each group and may be used to designate all contractions that yield the same algebraic term.
For example, in addition to the contraction shown in Eq.~\eqref{eq:example_elementary}, one should also consider the following contraction obtained by connecting $\cop{j}$ instead of $\cop{i}$:
\begin{equation}
\begin{split}
&\frac{1}{4}\sum_{ijk}^{\mathbb{O}} \sum_{abc}^{\mathbb{V}} \tens{f}{k}{c}\tens{t}{ab}{ij}
\contraction{\{}{\hat{a}}{{}_k \aop{c} \cop{a} \cop{b} }{a}
\no{{\color{sqred}\cop{k}} \aop{c} \cop{a} \cop{b} {\color{sqblue}\aop{j}} \aop{i}} \\
=& \frac{1}{4}\sum_{ijk}^{\mathbb{O}} \sum_{abc}^{\mathbb{V}} \tens{f}{k}{c}\tens{t}{ab}{ji}
\contraction{\{}{\hat{a}}{{}_k \aop{c} \cop{a} \cop{b} }{a}
\no{{\color{sqred}\cop{k}} \aop{c} \cop{a} \cop{b}  {\color{sqblue}\aop{i}}\aop{j}} \\
=& \frac{1}{4}\sum_{ijk}^{\mathbb{O}} \sum_{abc}^{\mathbb{V}} \tens{f}{k}{c}\tens{t}{ab}{ij}
\contraction{\{}{\hat{a}}{{}_k \aop{c} \cop{a} \cop{b} \aop{i} }{a}
\no{{\color{sqred}\cop{k}} \aop{c} \cop{a} \cop{b} \aop{j} {\color{sqblue}\aop{i}} } \\
\end{split}
\end{equation}
However, as shown above, since the tensor $\tens{t}{ab}{ij}$ is antisymmetric, after permuting the indices $i$ and $j$ and rearranging, this second contraction is identical to the one in Eq.~\eqref{eq:example_elementary}.
To account for these two equivalent contractions, when translating the incidence matrix in Eq.~\eqref{eq:example_elementary} to its algebraic form we multiply it by a combinatorial factor of 2.
This is an example of a Wick contraction that connects equivalent second-quantized operators, defined as operators of the same type (creation/annihilation) acting on the same orbital subspace.
The representation of composite contractions used in \wicked exploits this equivalence to minimize the number of terms generated and to facilitate the identification of identical terms.

A term resulting from Wick's theorem corresponds to a combination of elementary contractions, which we refer to as a composite contraction.
A composite contraction is represented by stacking rows of contraction matrices in the hypergraph incidence matrix.
Since an elementary contraction may appear more than once in a composite contraction, and the order of these is immaterial, the latter may be also represented with a multiset (e.g., $\{ \mathcal{C}_1, \mathcal{C}_1, \mathcal{C}_2 \}$).
The following example illustrates the representation of a pair of 2-leg contractions:
\begin{equation}
\frac{1}{4}\sum_{ijk}^{\mathbb{O}} \sum_{abc}^{\mathbb{V}} \tens{f}{k}{c}\tens{t}{ab}{ij}
\contraction{\{}{\hat{a}}{{}_k \aop{c} \cop{a} \cop{b} \aop{j}}{a}
\contraction[0.75em]{\{\cop{k}}{\hat{a}}{{}_{c} }{a}
\no{{\color{sqred}\cop{k}} {\color{sqorange}\aop{c}} {\color{sqblue}\cop{a}} \cop{b} \aop{j} {\color{sqgreen}\aop{i}}}
\leftrightarrow
\begin{array}{cc}
\begin{bmatrix}
0 & 0 \\
0 & {\color{sqorange}1}
\end{bmatrix} 
&
\begin{bmatrix}
0 & 0 \\
{\color{sqblue}1} & 0
\end{bmatrix}
\\[15pt]
\begin{bmatrix}
{\color{sqred}1} & 0 \\
0 & 0
\end{bmatrix} 
&
\begin{bmatrix}
0 & {\color{sqgreen}1} \\
0 & 0
\end{bmatrix}
\\[15pt]
\begin{bmatrix}
1& 0 \\
0 & 1
\end{bmatrix}
&
\begin{bmatrix}
0 & 2 \\
2 & 0
\end{bmatrix} \\
f & t
\end{array}.
\end{equation}
The top matrix row is a contraction that connects a pair of virtual annihilation/creation operators (orange/blue), while the middle matrix row is a contraction involving occupied creation/annihilation operator pair (red/green).
A more complex example is reported in Fig.~\ref{fig:hypergraph}, where in a CAS setting an elementary contraction of four operators yields a two-body cumulant (indicated in red).

Notice that the representation adopted here may be redundant if the only valid elementary contractions are those between operators acting on the same subspace.
However, there are cases when such an assumption is too restrictive.
For example, one way to generate spin integrated equations is to split the orbital subspaces into alpha and beta spin sets.
Then a term involving a mixed spin case cumulant would require expressing contractions of both alpha ($\mathbb{A}_\alpha$) and beta ($\mathbb{A}_\beta$) active orbitals, as shown in the following example:
\begin{equation}
\sum_{u_\alpha v_\alpha}^{\mathbb{A}_\alpha}
\sum_{u_\beta v_\beta}^{\mathbb{A}_\beta} \tens{f}{u_\alpha}{v_\alpha} \tens{t}{u_\beta}{v_\beta}
\no{
\underbrace{
\contraction{}{\hat{a}}{{}_{u_\alpha}\aop{v_\alpha} \cop{u_\beta}}{a}
\contraction{\cop{u_\alpha}}{\hat{a}}{{}_{v_\alpha}}{a}
\cop{u_\alpha}\aop{v_\alpha} \cop{u_\beta}\aop{v_\beta}
}_{\cumulant{u_\alpha u_\beta}{v_\alpha v_\beta}}
}
\leftrightarrow
\begin{array}{ccc}
\begin{bmatrix}
1 & 1 \\
0 & 0 \\
\end{bmatrix}
& 
\begin{bmatrix}
0 & 0 \\
1 & 1 \\
\end{bmatrix}
&
\\[15pt]
\begin{bmatrix}
1 & 1 \\
0 & 0 \\
\end{bmatrix}
&
\begin{bmatrix}
0 & 0 \\
1 & 1 \\
\end{bmatrix}
&
\begin{matrix}
\leftarrow \mathbb{A}_\alpha \\
\leftarrow \mathbb{A}_\beta
\end{matrix} \\
f & t
\end{array}.
\end{equation}

\subsection{Generation of Wick contractions}

\begin{figure}[h!]
   \centering
   \includegraphics{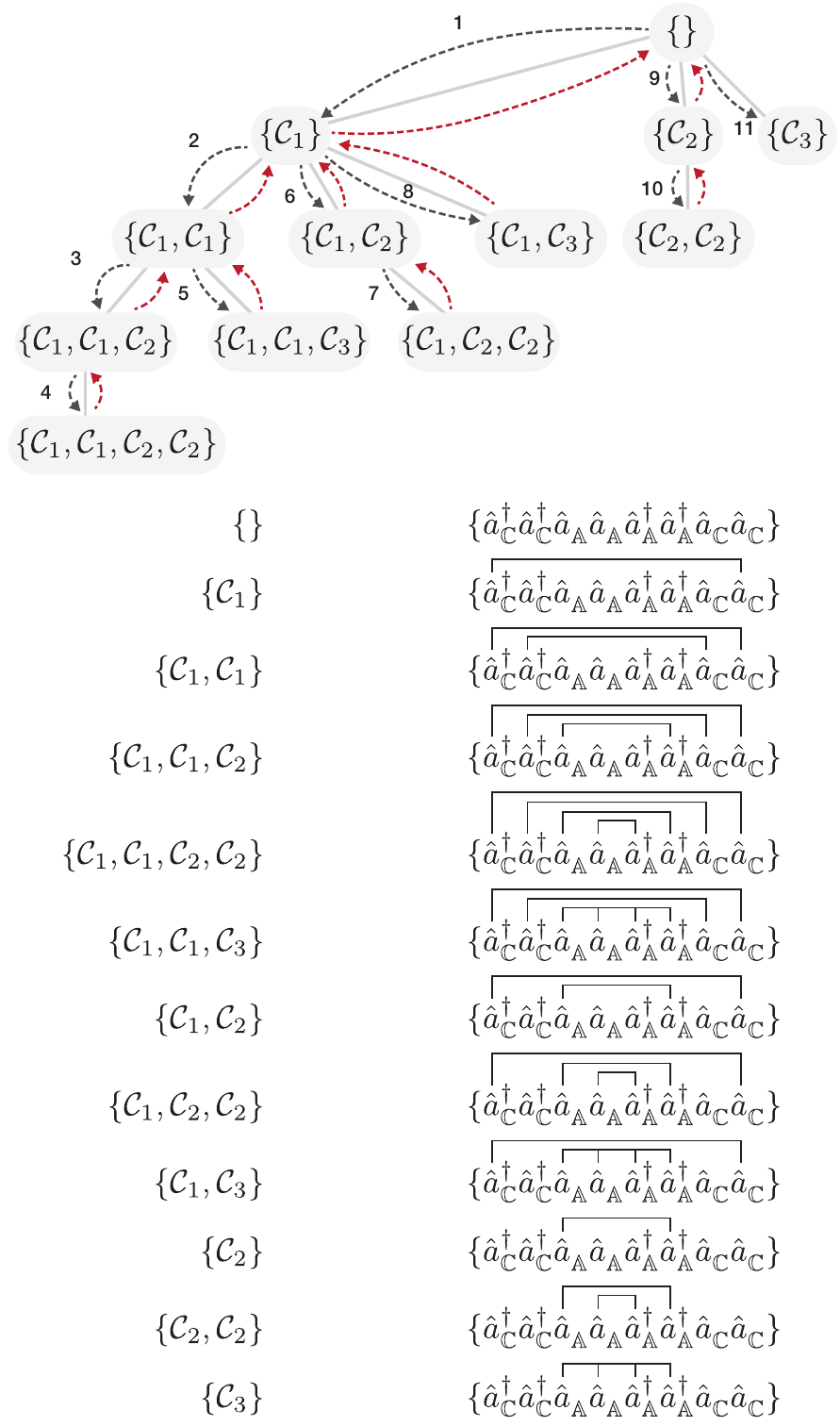}   \caption{Illustration of how the backtracking algorithm generates all Wick's theorem contractions for the term $\hat{V}_\mathbb{CCAA} \hat{T}_\mathbb{AACC}$ [see Eq.~\eqref{eq:example_term}]. Black dashed lines indicate steps in which elementary contractions are added to a composite contraction. Red dashed lines indicate backtracking steps.}
   \label{fig:backtracking}
\end{figure}

To identify all valid operator contractions efficiently, \wicked uses a backtracking algorithm.
The following subsections describes the steps of this algorithm.
\subsubsection{Generation of elementary contractions}
In the first step, the code identifies all the elementary contractions between the groups of second-quantized operators. We define these as all contractions of pairs of operators for occupied/unoccupied orbital subspaces ($\contraction{}{\hat{a}}{{}^{p}}{\hat{a}}\cop{p} \aop{q}$, $\contraction{}{\hat{a}}{{}_{q}}{\hat{a}}\aop{q} \cop{p}$) and the $2k$-leg contractions for general orbital subspaces [see Eq.~\eqref{eq:2cum}].
In defining elementary contractions, we consider only one out of the potentially many equivalent contractions that can be obtained by permuting operators of the same type (creation/annihilation) that act on the same orbital subspace.
For example, consider a CAS reference with orbitals partitioned as $\mathbb{S} = \mathbb{C} \cup \mathbb{A} \cup \mathbb{V}$, and the product $\hat{V}_\mathbb{CCAA} \hat{T}_\mathbb{AACC}$:
\begin{equation}
\label{eq:example_term}
\frac{1}{16} \sum_{c_1 c_2 c_3 c_4}^{\mathbb{C}}
\sum_{a_1 a_2 a_3 a_4}^{\mathbb{A}}
\tens{v}{c_3 c_4}{a_3 a_4} \tens{t}{a_1 a_2}{c_1 c_2} 
\no{ \cop{c_3} \cop{c_4} \aop{a_4} \aop{a_3} }
\no{ \cop{a_1} \cop{a_2} \aop{c_2} \aop{c_1} }
\end{equation}
If we ignore the orbital labels, this term may be written as $\no{ \copc \copc \aopa \aopa } \no{ \copa \copa \aopc \aopc }$.
There are three elementary contractions for this case. A contraction of a pair of core creation-annihilation operators, represented by the following edge in the incidence matrix
\begin{equation}
\contraction{\{}{\hat{a}}
{{}_{\mathbb{C}} \copc \aopa \aopa \no{} \copa \copa \aopa }
{\hat{a}}
\no{ \copc \copc \aopa \aopa }
\no{ \copa \copa \aopc \aopc } 
\leftrightarrow
\begin{bmatrix}
1 & 0 \\
0 & 0 \\
0 & 0
\end{bmatrix}
\begin{bmatrix}
0 & 1 \\
0 & 0 \\
0 & 0
\end{bmatrix}
= \mathcal{C}_1,
\end{equation}
a contraction between a pair of active annihilation-creation operators
\begin{equation}
\contraction{\{ \copc \copc}{\hat{a}}
{{}_{\mathbb{A}} \aopa \no{} \aopa}
{\hat{a}}
\no{ \copc \copc \aopa \aopa }
\no{ \copa \copa \aopc \aopc }
\leftrightarrow
\begin{bmatrix}
0 & 0 \\
0 & 1 \\
0 & 0
\end{bmatrix}
\begin{bmatrix}
0 & 0 \\
1 & 0 \\
0 & 0
\end{bmatrix}
= \mathcal{C}_2,
\end{equation}
and a 4-leg contraction among all the operators that act on the active orbitals
\begin{equation}
\contraction{\{ \copc \copc}{\hat{a}}
{{}_{\mathbb{A}} \aopa \no{} }
{\hat{a}}
\contraction{\{ \copc \copc \aopa}{\hat{a}}
{{}_{\mathbb{A}}  \no{} \copa }
{\hat{a}}
\no{ \copc \copc \aopa \aopa }
\no{ \copa \copa \aopc \aopc }
\leftrightarrow
\begin{bmatrix}
0 & 0 \\
0 & 2 \\
0 & 0
\end{bmatrix}
\begin{bmatrix}
0 & 0 \\
2 & 0 \\
0 & 0
\end{bmatrix}
= \mathcal{C}_3.
\end{equation}

\subsubsection{Generation of composite contractions by backtracking}
We employ a backtracking algorithm to identify all multisets that correspond to allowed combinations of elementary contractions.
This algorithm essentially visits all the relevant branches of a tree whose leaves represent all possible multisets of elementary contractions.
The backtracking algorithm used in \wicked is both efficient and flexible, since it applies to an arbitrary number of orbital subspaces.
As an example, Fig.~\ref{fig:backtracking} shows the steps taken by the backtracking algorithm to find all composite contractions that arise from the term in Eq.~\eqref{eq:example_term}.
We start from the fully uncontracted term, represented by the empty multiset $\{ \}$. Then in step (1), we test the solution obtained by adding the first elementary contraction ($\mathcal{C}_1$), and because this is a valid contraction, we add it to the current solution, obtaining the composite contraction $\{ \mathcal{C}_1 \}$.
In step (2), the current contraction is combined with the elementary contraction $\mathcal{C}_1$ again, leading to $\{ \mathcal{C}_1, \mathcal{C}_1 \}$, which is validated and added to the list of composite contractions.
At this point, it is no longer possible to add more $\mathcal{C}_1$ contractions, and we proceed by adding the next contraction in the list, $\mathcal{C}_2$. This contraction can be added up to two times, yielding the composite contractions $\{ \mathcal{C}_1, \mathcal{C}_1, \mathcal{C}_2 \}$ and $\{ \mathcal{C}_1, \mathcal{C}_1, \mathcal{C}_2, \mathcal{C}_2 \}$ [steps (3) and (4)].
At the end of step (4), all second-quantized operators are contracted, and the algorithm backtracks to $\{ \mathcal{C}_1, \mathcal{C}_1 \}$. In step (5) we test and add the contraction $\mathcal{C}_3$, and then backtrack up to the contraction $\{ \mathcal{C}_1 \}$.
The algorithm proceeds for six more steps, at which point all valid contractions have been enumerated.

Up to this stage we have selected only one of the possible permutations of contractions of equivalent set of operators (same type and orbitals space). Therefore, the backtracking algorithm produces a list of valid contractions that are equivalent to diagrams with distinct connectivity.
Nevertheless, it is still possible to generate isomorphic diagrams that yield equivalent algebraic terms, and these are dealt with in the next step.

\subsubsection{Contraction canonicalization}
An important task in automatic derivation of many-body equations is expressing equivalent terms into a canonical form so that they can be collected.
In \wicked, we employ early canonicalization, whereby contractions are canonicalized before converting them to algebraic expressions.
Currently, \wicked implements an exhaustive (combinatorial) canonicalization procedure of composite contractions.
Consider, for example, the commutator $[\hat{V}_\mathbb{AAAA},\hat{T}_\mathbb{AAAA}]$, where 
\begin{equation}
\hat{V}_\mathbb{AAAA} = 
\frac{1}{4} \sum_{stuv}^\mathbb{A} \tens{v}{st}{uv} \no{\sqop{st}{uv}}, \;
\hat{T}_\mathbb{AAAA} = 
\frac{1}{4} \sum_{wxyz}^\mathbb{A} \tens{t}{wx}{yz} \no{\sqop{wx}{yz}}.
\end{equation}
If we consider the product $\hat{V}_\mathbb{AAAA}\hat{T}_\mathbb{AAAA}$, one of the contributions to the fully contracted terms is (omitting the rows corresponding to the $\mathbb{C}$ and $\mathbb{V}$ spaces)
\begin{equation*}
\frac{1}{16}\tens{v}{st}{uv} \tens{t}{wx}{yz}  \no{
\contraction[0.75em]{}{\hat{a}}{{}_s \cop{t} \aop{v} \aop{u}\cop{w} \cop{x}\aop{z}}{a}
\contraction[0.75em]{\cop{s}}{\hat{a}}{{}_t \aop{v} \aop{u}\cop{w} \cop{x}}{a}
\contraction{\cop{s} \cop{t}}{\hat{a}}{{}_v  \aop{u}\cop{w} }{a}
\contraction{\cop{s} \cop{t} \aop{v}}{\hat{a}}{{}_u   }{a}
\cop{s} \cop{t} \aop{v} \aop{u}
\cop{w} \cop{x}\aop{z} \aop{y}
}
\leftrightarrow
\begin{array}{cc}
\begin{bmatrix}
2 & 0
\end{bmatrix}
&
\begin{bmatrix}
0 & 2
\end{bmatrix} \\[3pt]
\begin{bmatrix}
0 & 2
\end{bmatrix}
&
\begin{bmatrix}
2 & 0
\end{bmatrix} \\[3pt]
\begin{bmatrix}
2 & 2 \\
\end{bmatrix}
&
\begin{bmatrix}
2 & 2 \\
\end{bmatrix} \\
v & t
\end{array}.
\end{equation*}
The equivalent contraction for the product $\hat{T}_\mathbb{AAAA}\hat{V}_\mathbb{AAAA}$ is given by
\begin{equation*}
\frac{1}{16}\tens{v}{st}{uv} \tens{t}{wx}{yz}  \no{
\contraction[0.75em]{}{\hat{a}}{{}_w \cop{x} \aop{z} \aop{y}\cop{s} \cop{t}\aop{v}}{a}
\contraction[0.75em]{\cop{w}}{\hat{a}}{{}_x \aop{z} \aop{y}\cop{s} \cop{t}}{a}
\contraction{\cop{w} \cop{x}}{\hat{a}}{{}_z  \aop{y}\cop{s} }{a}
\contraction{\cop{w} \cop{x} \aop{z}}{\hat{a}}{{}_y   }{a}
\cop{w} \cop{x}\aop{z} \aop{y}
\cop{s} \cop{t} \aop{v} \aop{u}
}
\leftrightarrow
\begin{array}{cc}
\begin{bmatrix}
2 & 0
\end{bmatrix}
&
\begin{bmatrix}
0 & 2
\end{bmatrix} \\[3pt]
\begin{bmatrix}
0 & 2
\end{bmatrix}
&
\begin{bmatrix}
2 & 0
\end{bmatrix} \\[3pt]
\begin{bmatrix}
2 & 2 \\
\end{bmatrix}
&
\begin{bmatrix}
2 & 2 \\
\end{bmatrix} \\
t & v
\end{array}.
\end{equation*}
These two terms yield identical algebraic expressions and this can be shown by expressing both of them in canonical form.
In manipulating the incidence matrix we are allowed to reorder the contractions (rows) and operators (columns) as long as the resulting expression is equivalent to the original contraction.
In the present example, starting from the contraction of  $\hat{T}_\mathbb{AAAA}\hat{V}_\mathbb{AAAA}$, we may permute the two groups of operators (by swapping columns):
\begin{equation*}
\frac{1}{16}\tens{v}{st}{uv} \tens{t}{wx}{yz}  \no{
\contraction{}{\hat{a}}{{}_s \cop{t} \aop{v} \aop{u}\cop{w} \cop{x}\aop{z}}{a}
\contraction{\cop{s}}{\hat{a}}{{}_t \aop{v} \aop{u}\cop{w} \cop{x}}{a}
\contraction[0.75em]{\cop{s} \cop{t}}{\hat{a}}{{}_v  \aop{u}\cop{w} }{a}
\contraction[0.75em]{\cop{s} \cop{t} \aop{v}}{\hat{a}}{{}_u   }{a}
\cop{s} \cop{t} \aop{v} \aop{u}
\cop{w} \cop{x}\aop{z} \aop{y}
}
\leftrightarrow
\begin{array}{cc}
\begin{bmatrix}
0 & 2
\end{bmatrix}
&
\begin{bmatrix}
2 & 0
\end{bmatrix} \\[3pt]
\begin{bmatrix}
2 & 0
\end{bmatrix}
&
\begin{bmatrix}
0 & 2
\end{bmatrix} \\[3pt]
\begin{bmatrix}
2 & 2 \\
\end{bmatrix}
&
\begin{bmatrix}
2 & 2 \\
\end{bmatrix} \\
v & t
\end{array}.
\end{equation*}
and then interchange the order of contractions (by swapping the top two rows) to obtain the same incidence matrix for the corresponding term arising from $\hat{V}_\mathbb{AAAA}\hat{T}_\mathbb{AAAA}$.
Once expressed in a canonical form, these two contributions cancel, giving a zero contribution for the term $\braket{\mref | [\hat{V}_\mathbb{AAAA},\hat{T}_\mathbb{AAAA}] | \mref}$.

To automate the identification of equivalent terms, we define an ordering of the directed hypergraph incidence matrices, and we define the canonical form as the minimal element out of all possible incidence matrices that represent the same term.
The details of this procedure are discussed in Appendix~\ref{app:matrix_ordering}.

\subsubsection{Conversion of contractions to algebraic expressions}
After all composite contractions are generated and equivalent contributions are combined, each term is converted to an algebraic expression.
Consider, for example, the following contraction
\begin{equation}
\label{eq:c1_c1_example}
\{ \mathcal{C}_1, \mathcal{C}_1 \} \rightarrow 
\contraction[0.75 em]{\{}{\hat{a}}
{{}_{\mathbb{C}} \copc \aopa \aopa  \copa \copa \aopc }
{\hat{a}}
\contraction{\{ \copc}{\hat{a}}
{{}_{\mathbb{C}} \aopa \aopa  \copa \copa }
{\hat{a}}
\no{ \copc \copc \aopa \aopa \copa \copa \aopc \aopc }.
\end{equation}
To convert it to an algebraic expression, we first assign distinct indices to the operators in the order in which they appear in the second-quantized operators
\begin{equation}
\frac{1}{16} \tens{v}{c_1 c_2}{a_2 a_1} \tens{t}{a_3 a_4}{c_4 c_3}
\contraction[0.75 em]{\{}{\hat{a}}
{{}_{c_1} \cop{c_2} \aop{a_1} \aop{a_2} \cop{a_3} \cop{a_1} \aop{c_3} }
{\hat{a}}
\contraction{\{\cop{c_1}}{\hat{a}}
{{}_{c_2} \aop{a_1} \aop{a_2} \cop{a_3} \cop{a_1} }
{\hat{a}}
\no{ \cop{c_1} \cop{c_2} \aop{a_1} \aop{a_2} \cop{a_3} \cop{a_4} \aop{c_3} \aop{c_4} }.
\end{equation}
Next, we reorder this term so that second-quantized operators that appear in the same contraction are adjacent, keeping track of sign factors that arise from permutations
\begin{equation}
\frac{1}{16} \tens{v}{c_1 c_2}{a_2 a_1} \tens{t}{a_3 a_4}{c_4 c_3}
\underbrace{
\contraction{}{\hat{a}}{{}_{c_1}}{\hat{a}}
\cop{c_1}\aop{c_4}
}_{\delta_{c_1 c_4}}
\underbrace{
\contraction{}{\hat{a}}{{}_{c_2}}{\hat{a}}
\cop{c_2}\aop{c_3}
}_{\delta_{c_2 c_3}}
\no{\aop{a_1} \aop{a_2} \cop{a_3} \cop{a_4}}.
\end{equation}
After elimination of the Kronecker delta factors we arrive at the expression
\begin{equation}
\frac{1}{16} \tens{v}{c_1 c_2}{a_2 a_1} \tens{t}{a_3 a_4}{c_1 c_2}
\no{\aop{a_1} \aop{a_2} \cop{a_3} \cop{a_4}},
\end{equation}
which is brought into a canonical form by rearranging the second-quantized operators and relabeling the indices.
In the last step, this contribution is multiplied by a combinatorial factor (2) that keeps into account the following identical contribution that differs by a permutations of two equivalent operators
\begin{equation}
\{ \mathcal{C}_1, \mathcal{C}_1 \} \rightarrow 
\contraction[0.75 em]{\{}{\hat{a}}
{{}_{\mathbb{C}} \copc \aopa \aopa  \copa \copa }
{\hat{a}}
\contraction{\{ \copc}{\hat{a}}
{{}_{\mathbb{C}} \aopa \aopa  \copa \copa \aopc}
{\hat{a}}
\no{ \copc \copc \aopa \aopa \copa \copa \aopc \aopc }.
\end{equation}
The resulting canonical expression for the contraction $\{ \mathcal{C}_1, \mathcal{C}_1 \}$ is:
\begin{equation}
\frac{1}{8} \sum_{c_1 c_2}^{\mathbb{C}} \sum_{a_1 a_2 a_3 a_4}^{\mathbb{A}}  \tens{v}{c_1 c_2}{a_3 a_4} \tens{t}{a_1 a_2}{c_1 c_2}
\no{\cop{a_1} \cop{a_2} \aop{a_4} \aop{a_3} }.
\end{equation}
The equation for the combinatorial factor of a general contraction is reported in Appendix~\ref{app:combinatorial_factor}.
The results of Wick's theorem lead to a sum of normal-ordered operators that may be subsequently processed or implemented as tensor contractions.

\subsection{Implementation}
The algebra of second-quantized operators and the backtracking algorithm to generate Wick contractions are implemented in \wicked, an open-source C++ library exposed as a Python module via the \textsc{pybind11} library.\cite{jakob2017pybind11}
\wicked's C++ library defines various classes used to represent diagrams (using the hypergraph incidence matrix) and algebraic terms that represent equations in terms of explicit orbital indices.
The \wicked repository includes several Jupyter-notebook tutorials on the use of the API and various examples applications (including the ones described in this paper).
Currently, \wicked supports the derivation of spinorbital or spin-integrated expressions.
Spin adaptation of these equations is an extension planned for future releases of the code.

\section{Example applications}
\label{sec:examples}

In this section we showcase two applications of \wicked to the derivation of single-reference and multireference many-body theories.

\subsection{High-order coupled cluster theory}

In our first example we derive expressions for the coupled cluster (CC) theory residuals ($\tens{r}{ab\cdots}{ij\cdots}$):
\begin{equation}
\label{eq:ccres}
\tens{r}{ab\cdots}{ij\cdots} = \braket{\Phi| \no{\sqop{ij\cdots}{ab\cdots}} \bar{H} | \Phi} 
\end{equation}
where $\bar{H} = \exp(-\hat{T}) \hat{H} \exp(\hat{T})$ is the similarity-transformed Hamiltonian.
Here the operator $\hat{T} = \hat{T}_1 + \hat{T}_2 + \ldots + \hat{T}_n$ is a sum of particle-hole excitation operators up to order $n$, with a generic operator $\hat{T}_k$ defined as
\begin{equation}
\hat{T}_k = \frac{1}{(k!)^2} \sum_{ij\cdots}^\mathbb{O} \sum_{ab\cdots}^\mathbb{V} \tens{t}{ab\cdots} {ij\cdots}\no{\sqop{ab\cdots}{ij\cdots}}.
\end{equation}
Notice that our index notation for the cluster amplitudes $\tens{t}{ab\cdots} {ij\cdots}$ corresponds to the traditional notation\cite{Crawford.2000.10.1002/9780470125915.ch2,Shavitt.2009} with upper/lower indices swapped.
We obtain the CC amplitude expressions by first evaluating the many-body operator $\bar{H}$
\begin{equation}
\bar{H} = E_0 + \sum_{pq} \tens{\bar{H}}{p}{q} \no{\sqop{p}{q}}
+ \frac{1}{4} \sum_{pqrs} \tens{\bar{H}}{pq}{rs} \no{\sqop{pq}{rs}} + \ldots,
\end{equation}
and then by extracting the tensor components $\tens{\bar{H}}{ab\cdots}{ij\cdots}$ corresponding to the particle-hole excitation operators $\no{\sqop{ab\cdots}{ij\cdots}}$.
From Wick's theorem it follows that the $k$-th order CC residuals $\braket{\Phi| \no{\sqop{ij\cdots}{ab\cdots}} \bar{H} | \Phi} $ are related to the tensor elements $\tens{\bar{H}}{ab\cdots}{ij\cdots}$ by separate antisymmetrization of the upper and lower indices:
\begin{equation}
\tens{r}{ab\cdots}{ij\cdots} =(k!)^2 \mathcal{A}_{ij\cdots} \mathcal{A}_{ab\cdots}\tens{\bar{H}}{ab\cdots}{ij\cdots},
\label{eq:residual_Hbar}
\end{equation}
where $\mathcal{A}_{pq\cdots}$ is the antisymmetrizer, defined as the sum over the $k!$ permutations of the indices $p,q,\ldots$ divided by $1/k!$.
Note that Eq.~\eqref{eq:residual_Hbar} applies only in the case of single-reference theories, and in the multireference case the residual is a more complicated expression of the tensor $\tens{\bar{H}}{ab\cdots}{ij\cdots}$, density matrices, and cumulants.

The number of unique terms (diagrams) in CC with excitation level $n$ ranging from 2 (CCSD) to 8 (CCSDTQPH78) computed with \wicked are reported in Tab.~\ref{tab:ccn}.
At each truncation level, we confirmed that \wicked yields the same number of diagrams reported by Kállay and Surján in their study of arbitrary order CC methods.\cite{Kallay.2001.10.1063/1.1383290}
Together with the number of diagrams, we also report the runtime of \wicked, showing that derivation of even CC equations with up to octuple excitations can be performed in less than a minute.

As a final test, we numerically validated the CCSD and CCSDT residual equations derived by \wicked by implementing them in a pilot Python code. We used integrals computed with Psi4\cite{Smith.2020.10.1063/5.0006002} and generated code that calls the tensor contraction function \textsc{einsum} implemented in the \textsc{numpy} library\cite{Harris.2020.10.1038/s41586-020-2649-2} and verified that this implementation matches reference energies.

\begin{table}[htbp]
   \caption{Evaluation of the coupled cluster residual equations with \wicked. For each level of theory, the table reports the execution time (on an Apple M1 Max laptop) and the number of unique terms contributing to the residual equations at a given particle-hole excitation level.}
   \centering
   \begin{tabular}{@{} lccccccccccc @{}}
        	\hline
	
	\hline
   	Theory  & Time  & \multicolumn{9}{c}{Diagrams per excitation level} \\
	  &  (s) & 0 & 1 & 2 & 3 & 4 & 5 & 6 & 7  & 8 \\
	\hline
	CCSD    & 0.1   & 3 & 14 & 31 \\		
	CCSDT   & 0.7  & 3 & 15 & 37 & 47 \\	
	CCSDTQ  & 2.4  & 3 & 15 & 38 & 53 & 74 \\
	CCSDTQP & 6.3 & 3 & 15 & 38 & 54 & 80 & 99 \\	
	CCSDTQPH & 13.8 & 3 & 15 & 38 & 54 & 81 & 105 & 135\\
	CCSDTQPH7 & 26.0 & 3 & 15 & 38 & 54 & 81 & 106 & 141 & 169 \\
	CCSDTQPH78 & 45.4 & 3 & 15 & 38 & 54 & 81 & 106 & 142 & 175 & 215 \\
        	\hline
	
	\hline	
   \end{tabular}
   \label{tab:ccn}
\end{table}

\subsection{Second-order driven similarity renormalization group multireference perturbation theory}

\begin{lstlisting}[language=Python,  float=*, caption={\wicked code to derive the DSRG-MRPT2 energy expression reported in Eq.~\eqref{eq:dsrg_second_order_energy_terms}.}, label={lst:dsrg-mrpt2}]
import wicked as w

# define orbital subspaces
w.add_space("c", "fermion", "occupied", ["i", "j","k"])
w.add_space("a", "fermion", "general", ["u","v","w","x","y","z"])
w.add_space("v", "fermion", "unoccupied", ["a", "b","c"])

# define the zeroth - and first -order Hamiltonian
H0 = w.op("F",["c+ c","a+ a", "v+ v"])
F1 = w.utils.gen_op("F",1,"cav","cav",diagonal=False)
V1 = w.utils.gen_op("V",2,"cav","cav")
H1 = F1 + V1

# define the cluster operator
T1 = w.utils.gen_op("T1",1,"av","ca",diagonal=False)
T2 = w.utils.gen_op("T2",2,"av","ca",diagonal=False)
A = T1 - T1.adjoint() + T2 - T2.adjoint()

# define the effective first -order operator
Hbar1 = H1 + w.commutator(H0,A)

# define the second -order energy
E2 = w.commutator(H1,A) + w.rational(1,2) * w.commutator(H0,A,A)

# generate expressions
wt = w.WickTheorem()
Hbar1expr = wt.contract(Hbar1 , 1, 1) + wt.contract(Hbar1 , 2, 2)
E2expr = wt.contract(E2, 0, 0)

print(E2expr)
\end{lstlisting}

The second example uses \wicked to implement the second-order driven similarity renormalization multireference perturbation theory (DSRG-MRPT2).\cite{Li.2015.10.1021/acs.jctc.5b00134}
For the sake of brevity, here we focus on the essential aspects of DSRG-MRPT2 relevant to the derivation of the corresponding equations and direct the interested reader to consult a recent review for further details.\cite{Li.2019.10.1146/annurev-physchem-042018-052416}
In the unitary DSRG formalism, the bare Hamiltonian ($\hat{H}$) is diagonalized by a continuous unitary transformation:
\begin{equation}
\label{eq:dsrg_transformation}
    \hat{H} \mapsto \bar{H}(s) = e^{-\hat{A} (s)} \hat{H} e^{\hat{A} (s)},
\end{equation}
where $\bar{H}(s)$ is the transformed Hamiltonian and $\hat{A} (s) = \hat{T} (s) - \hat{T}^\dagger (s)$ is an anti-Hermitian operator that depends on a time-like parameter $s$ defined in the range $[0, \infty)$.
The DSRG energy is given by the expectation value of $\bar{H}(s)$ with respect to a CASSCF reference state $\Psi_0$, $E(s) = \braket{\Psi_0 | \bar{H}(s) | \Psi_0}$. The operator $\hat{A} (s)$ is obtained by solving a set of Fock-space many-body equations of the form $[\bar{H} (s)]_\text{N} = \hat{R} (s)$, where the subscript ``N'' indicates the non-diagonal part of $\bar{H} (s)$. The tensor component of the source operator $\hat{R} (s)$ takes the form
\begin{equation}
\label{eq:source_operator}
\tens{r}{ab\cdots}{ij\cdots}(s) = [\tens{\bar{H}}{ab\cdots}{ij\cdots}(s) + \tens{t}{ab\cdots}{ij\cdots}(s) \Delta_{ab\cdots}^{ij\cdots}] e^{-s(\Delta_{ab\cdots}^{ij\cdots})^2},
\end{equation}
and $\tens{r}{ij\cdots}{ab\cdots}(s) = [\tens{r}{ab\cdots}{ij\cdots}(s)]^*$.  In Eq.~\eqref{eq:source_operator} $\Delta_{ab\cdots}^{ij\cdots}$ is a generalized M{\o}ller--Plesset denominator defined in terms of the diagonal components of the Fock matrix ($\dfock{p} = \tens{f}{p}{p}$),
\begin{align}
    \Delta_{ab\cdots}^{ij\cdots} &= \dfock{i} + \dfock{j} + \cdots - \dfock{a} - \dfock{b} - \cdots.
\end{align}
The DSRG-MRPT2 uses a diagonal Fock partitioning, whereby the Hamiltonian normal-ordered with respect to $\mref$ is split according to $\hat{H} = \hat{H}^{(0)} + \xi \hat{H}^{(1)}$. The zeroth-order Hamiltonian $\hat{H}^{(0)}$ is the sum of the reference energy ($E_{0}$) and a diagonal one-body operator [$\hat{F}^{(0)}$]:
\begin{equation}
    \hat{H}^{(0)} = E_{0} + \hat{F}^{(0)} =E_{0} + \sum_{p}^{\mathbb{G}} \epsilon_{p} \no{\sqop{p}{p}}.
\end{equation}
The first-order amplitudes that enter the operator $\hat{A}^{(1)}(s)$ are obtained by solving the following linear equation (omitting the variable ``$s$'')
\begin{equation}
\label{eq:dsrg_first_order_amps}
[\bar{H}^{(1)}]_\text{N} \equiv \left( \hat{H}^{(1)} + [ \hat{H}^{(0)}, \hat{A}^{(1)} ] \right)_\text{N} = \hat{R}^{(1)},
\end{equation}
while the second-order energy is given by
\begin{equation}
\label{eq:dsrg_second_order_energy}
E^{(2)} =
\braket{\mref | [\hat{H}^{(1)} ,\hat{A}^{(1)} ] +\frac{1}{2}  [[\hat{H}^{(0)} ,\hat{A}^{(1)}],\hat{A}^{(1)}] | \mref}.
\end{equation}

Listing~\ref{lst:dsrg-mrpt2} shows how to use \wicked to evaluate Eqs.~\eqref{eq:dsrg_first_order_amps} and \eqref{eq:dsrg_second_order_energy} in terms of molecular integrals and the first-order amplitude equations.
In this listing we define the $\mathbb{C}$, $\mathbb{A}$, and $\mathbb{V}$ orbital subspaces (lines 4--6), specify the form of the operators $\hat{H}^{(0)}$, $\hat{H}^{(1)}$ (lines 9--12) and the anti-Hermitian operator $\hat{A}^{(1)}$ (lines 15--17), form the operator $\bar{H}^{(1)}$ (line 20), define the second-order energy (line 23), and finally apply Wick's theorem to evaluate $\bar{H}^{(1)}$ and the second-order energy (lines 27 and 28).
The resulting second-order energy expression contains 226 contractions.
This number may be reduced to 24 contractions by introducing the following intermediate $\tilde{H}^{(1)} = \hat{H}^{(1)}  + \hat{R}$, which allows us to rewrite $E^{(2)}$ as\cite{Li.2015.10.1021/acs.jctc.5b00134}
\begin{equation}
\label{eq:dsrg_second_order_energy2}
E^{(2)} =\braket{\mref | [\tilde{H}^{(1)} ,\hat{T}^{(1)} ] | \mref},
\end{equation}
where $\tilde{H}^{(1)}$ is a general two-body operator
\begin{equation}
\tilde{H}^{(1)} = E_0 + \sum_{pq} \tens{\tilde{f}}{p}{q} \no{\sqop{p}{q}}
+ \frac{1}{4} \sum_{pqrs} \tens{\tilde{v}}{pq}{rs} \no{\sqop{pq}{rs}}
\end{equation}

The energy expression derived by \wicked using Eq.~\eqref{eq:dsrg_second_order_energy2} is reproduced below
\begin{equation}
\label{eq:dsrg_second_order_energy_terms}
\begin{split}
E^{(2)} =& {\eta}^{v}_{u} {\tilde{f}}^{u}_{i} {t}^{i}_{v} + {\tilde{f}}^{a}_{i} {t}^{i}_{a} + {\tilde{f}}^{a}_{u} {\gamma}^{u}_{v} {t}^{v}_{a} \\
&-\frac{1}{2} {\tilde{f}}^{u}_{i} {\lambda}^{w x}_{u v} {t}^{i v}_{w x}  
-\frac{1}{2} {\tilde{f}}^{a}_{u} {\lambda}^{u x}_{v w} {t}^{v w}_{x a} \\
&+\frac{1}{2} {\lambda}^{w x}_{u v} {t}^{i}_{x} {\tilde{v}}^{u v}_{i w}  
-\frac{1}{2} {\lambda}^{w x}_{u v} {t}^{u}_{a} {\tilde{v}}^{v a}_{w x} \\
&+\frac{1}{2} {\eta}^{v}_{u} {\eta}^{x}_{w} {\gamma}^{z}_{y} {t}^{i y}_{v x} {\tilde{v}}^{u w}_{i z} 
+\frac{1}{4} {\eta}^{v}_{u} {\eta}^{x}_{w} {t}^{i j}_{v x} {\tilde{v}}^{u w}_{i j} 
+\frac{1}{2} {\eta}^{v}_{u} {\gamma}^{x}_{w} {\gamma}^{z}_{y} {t}^{w y}_{v a} {\tilde{v}}^{u a}_{x z} \\ 
&+ {\eta}^{v}_{u} {\gamma}^{x}_{w} {t}^{i w}_{v a} {\tilde{v}}^{u a}_{i x}
- {\eta}^{v}_{u} {\lambda}^{y z}_{w x} {t}^{i w}_{v z} {\tilde{v}}^{u x}_{i y}
+\frac{1}{4} {\eta}^{v}_{u} {\lambda}^{y z}_{w x} {t}^{w x}_{v a} {\tilde{v}}^{u a}_{y z} \\ 
&+\frac{1}{2} {\eta}^{v}_{u} {t}^{i j}_{v a} {\tilde{v}}^{u a}_{i j} 
+\frac{1}{4} {\gamma}^{v}_{u} {\gamma}^{x}_{w} {t}^{u w}_{a b} {\tilde{v}}^{a b}_{v x} 
+\frac{1}{4} {\gamma}^{v}_{u} {\lambda}^{y z}_{w x} {t}^{i u}_{y z} {\tilde{v}}^{w x}_{i v} \\ 
&- {\gamma}^{v}_{u} {\lambda}^{y z}_{w x} {t}^{u w}_{z a} {\tilde{v}}^{x a}_{v y} 
+\frac{1}{2} {\gamma}^{v}_{u} {t}^{i u}_{a b} {\tilde{v}}^{a b}_{i v} 
+\frac{1}{8} {\lambda}^{w x}_{u v} {t}^{i j}_{w x} {\tilde{v}}^{u v}_{i j} \\ 
&- {\lambda}^{w x}_{u v} {t}^{i u}_{x a} {\tilde{v}}^{v a}_{i w} 
+\frac{1}{8} {\lambda}^{w x}_{u v} {t}^{u v}_{a b} {\tilde{v}}^{a b}_{w x} 
+\frac{1}{4} {\lambda}^{x y z}_{u v w} {t}^{i u}_{y z} {\tilde{v}}^{v w}_{i x} \\ 
&-\frac{1}{4} {\lambda}^{x y z}_{u v w} {t}^{u v}_{z a} {\tilde{v}}^{w a}_{x y} 
+\frac{1}{4} {t}^{i j}_{a b} {\tilde{v}}^{a b}_{i j}
\end{split}
\end{equation}
In this expression we use Einstein notation and assign the labels to the orbital subspaces in the following way: $i,j \in \mathbb{C}$, $u,v,w,x,y,z \in \mathbb{A}$, and $a,b \in \mathbb{C}$.
We numerically validated both versions of the DSRG-MRPT2 equations [Eqs.~\eqref{eq:dsrg_second_order_energy} and \eqref{eq:dsrg_second_order_energy2}] by implementing them in a pilot Python code using an interface to \textsc{Forte}.\cite{forte}
This implementation also used \wicked to derive and test equations for the first-order amplitude equations.

\section{Discussion and current limitations}
\label{sec:discussion}

We have presented a computational strategy to evaluate Mukherjee and Kutzelnigg's generalized version of Wick's theorem in the case of an arbitrary number of orbital subspaces.
Our approach represents Wick contractions using directed hypergraphs represented as incidence matrices.
To rapidly evaluate all Wick terms, it enumerates all elementary contractions among the operators to be contracted and uses a backtracking algorithm to generate combinations of elementary contractions (composite contractions).
This algorithm forms the basis for the automated derivation of many-body equations implemented in the open-source software library \wicked .
We have illustrated the utility of this code by deriving the single-reference coupled cluster equations through octuple excitations and the energy expressions for the second-order multireference driven similarity renormalization group perturbation theory (DSRG-MRPT2).
The correctness of these equations was verified via pilot numerical implementations of CCSD, CCSDT, and DSRG-MRPT2.

In its current form, \wicked is capable of deriving many-body equations that can be used for the purpose of exploring new theories and developing pilot implementations.
In the case of multireference theories, \wicked is particularly useful for the derivation of equations of internally-contracted methods.
Although in our examples we have emphasized the formalism for a general reference state, our framework is equally applicable to derive equations for state-averaged theories of multiple electronic states\cite{Aoto.2016.10.1063/1.4941604b,Li.2018.10.1063/1.5019793} and, more generally, ensembles of states.
Given its ability to describe an arbitrary number of fermionic spaces, another unexplored application of \wicked would be the derivation of many-body equations for systems involving electrons and fermionic nuclei.\cite{Pavosevic.2018.10.1021/acs.jctc.8b01120,Hammes-Schiffer.2021.10.1063/5.0053576}

In the future, we plan to expand the capabilities of \wicked in several directions.
To generate efficient implementations, it would be desirable to develop post-processing modules to spin adapt the equations generated by \wicked, and to identify reusable tensor intermediates.
It would also be interesting to extend the Wick's theorem kernel to more general references.
Examples include the ability to consider number-symmetry-broken vacua that require the inclusion of anomalous contractions (e.g., among operators of the same type like $\contraction{}{\hat{a}}{{}_{i}}{\hat{a}}\cop{i} \cop{j}$).
Similarly, it would be desirable to expand \wicked to mixed fermionic/bosonic fields to enable the derivation of many-body equations for strongly interacting electron-phonon\cite{White.2020.10.1063/5.0033132} and electron-photon systems.\cite{Haugland.2020.10.1103/PhysRevX.10.041043,DePrince.2021.10.1063/5.0038748}

\begin{acknowledgements}	
The author would like to thank Jonathon Misiewicz and Ilias Magoulas for providing feedback on a draft of this work.
Thanks also go to Matthias Hanauer, Andreas K\"{o}hn, Liguo Kong, Chenyang Li, and Luther Blissett for insightful discussions in the early stages of conceptualizing the design of \wicked.
This work was supported by a grant from the U.S. Department of Energy under Award No. DE-SC0016004 and a Camille Dreyfus Teacher-Scholar Award (TC-18-045).
\end{acknowledgements}	

\appendix

\section{Ordering of incidence matrices and directed hypergraph canonical form}
\label{app:matrix_ordering}

In this appendix, we define an ordering on the set of all possible equivalent incidence matrices representing a Wick contraction of $K$ operators and $L$ elementary contractions.
We then discuss how this ordering is used to define a canonical form of the directed hypergraph representation.
An incidence matrix $\mathcal{W}$ may be represented by a table of matrices and labels:
\begin{equation}
\label{app:eq:incidence}
\mathcal{W} = 
\begin{array}{cccc}
\mathbf{C}_{1L} & \mathbf{C}_{2L} & \cdots & \mathbf{C}_{KL} \\
\vdots & \vdots & \vdots \\
\mathbf{C}_{12} & \mathbf{C}_{22} & \cdots & \mathbf{C}_{K2}\\
\mathbf{C}_{11} & \mathbf{C}_{21} & \cdots & \mathbf{C}_{K1}\\
\mathbf{N}_{1} & \mathbf{N}_{2} & \cdots & \mathbf{N}_{K} \\
\omega_1 & \omega_2 & \cdots & \omega_K
\end{array},
\end{equation}
where $\mathbf{N}_{i}$ and $\omega_i$ are the operator matrix and label corresponding to the $i$-th operator, respectively, while $\mathbf{C}_{ij}$ is the contraction matrix for the $i$-th operator and $j$-th elementary contraction.

Now consider an equivalent incidence matrix $\mathcal{W}'$ obtained by permuting the rows and columns of the incidence matrix shown in Eq.~\eqref{app:eq:incidence}, which we write as
\begin{equation}
\label{app:eq:incidence2}
\mathcal{W}' = 
\begin{array}{cccc}
\mathbf{C}'_{1L} & \mathbf{C}'_{2L} & \cdots & \mathbf{C}'_{KL} \\
\vdots & \vdots & \vdots \\
\mathbf{C}'_{12} & \mathbf{C}'_{22} & \cdots & \mathbf{C}'_{K2}\\
\mathbf{C}'_{11} & \mathbf{C}'_{21} & \cdots & \mathbf{C}'_{K1}\\
\mathbf{N}'_{1} & \mathbf{N}'_{2} & \cdots & \mathbf{N}'_{K} \\
\omega'_1 & \omega'_2 & \cdots & \omega'_K
\end{array}
\end{equation}

For two given incidence matrices $\mathcal{W}$ and $\mathcal{W}'$, we test for the condition $\mathcal{W} < \mathcal{W}'$ using lexicographic ordering.
Specifically, we first compare the operator labels and matrices and identify the first place $i$ where the elements $(\mathbf{N}_i,\omega_i)$ and $(\mathbf{N}'_i,\omega'_i)$ differ.
If $\omega_i < \omega'_i$, then $\mathcal{W} < \mathcal{W}'$.
If $\omega_i = \omega'_i$, then we test for $\mathbf{N}_i < \mathbf{N}'_i$, by comparing the entries of the two columns of $\mathbf{N}_i$ and $\mathbf{N}'_i$ lexicographically.
If $\omega_i > \omega'_i$ then $\mathcal{W}' < \mathcal{W}$.
If the operator matrices and labels are identical, we then proceed to compare the contraction matrices $\mathbf{C}_{ij}$ and $\mathbf{C}'_{ij}$, one row at a time. The first place where two contraction matrices differ, we compare $\mathbf{C}_{ij}$ and $\mathbf{C}'_{ij}$ lexicographically. If $\mathbf{C}_{ij} < \mathbf{C}'_{ij}$ then $\mathcal{W} < \mathcal{W}'$, otherwise $\mathcal{W}' < \mathcal{W}$.

Once an ordering of the incidence matrices is defined, we consider the subset of valid incidence matrices out of the $K! L!$ possible ones obtained by permuting the position of the $K$ operators and $L$ elementary contractions. The element of this subset $\mathcal{W}$ that is minimal (i.e., $\mathcal{W} < \mathcal{W}'$ for each $\mathcal{W}'$) is the Wick contraction canonical form.

\section{Combinatorial factor}
\label{app:combinatorial_factor}

In this appendix, we report the equations for the combinatorial factor associated with the incidence matrix representation of a Wick contraction.
To illustrate how the combinatorial factor is derived, we use a simple example.
Consider the following three-leg (red) and two-leg (blue) contractions of a group of $n$ second-quantized operators:
\begin{equation}
\no{
\cdots
{\color{sqred}
\lcontraction{}{\hat{a}}{\quad\quad\quad}{a}
\lcontraction{aaa}{\hat{a}}{\quad\quad\quad}{a}
\lcontraction{aa}{\hat{a}}{\quad\quad\quad}{a}
}
\qop{1}\cdots \qop{n}
\cdots
}
\rightarrow
\no{
\cdots
{\color{sqred}
\lcontraction{}{\hat{a}}{\quad\quad\quad}{a}
\lcontraction{aaa}{\hat{a}}{\quad\quad\quad}{a}
\lcontraction{aa}{\hat{a}}{\quad\quad\quad}{a}
}
{\color{sqblue}
\lcontraction[0.75em]{a}{\hat{a}}{\quad\quad \,}{a}
\lcontraction[0.75em]{aqqqq}{\hat{a}}{\quad\quad \,}{a}
}
\qop{1}\cdots \qop{n}
\cdots
}.
\end{equation}
We assume that the operators $\qop{1}\cdots \qop{n}$ are of the same type (creation or annihilation) and act on the same orbital subspace.
The number of permutations of the contractions legs is given by number of ways one can assign the three legs of the red contraction ($n$ choose 3) times the number of ways one can assign the two legs of the blue contraction ($n-3$ choose 2) to the remaining $n-3$ uncontracted operators.
It is easy to see that this quantity corresponds to a multinomial factor
\begin{equation}
\binom{n}{3} \binom{n-3}{2}
= \frac{n!}{3! 2! (n-3- 2)!}
= \binom{n}{2,3,n-2-3},
\label{eq:app:example}
\end{equation}
where the multinomial factor is defined as
\begin{equation}
\binom{n}{c_1,\ldots,c_m} = \frac{n!}{c_1! \cdots c_m!}.
\end{equation}
The total combinatorial factor for a Wick contraction is the product of combinatorial factors for each type of operator (creation/annihilation) and orbital subspace.
In the case of a Wick contraction involving two or more equivalent contractions, an additional numerical factor must be included to avoid double counting contractions that are indistinguishable.
The following example shows the case of a composite contraction with three equivalent elementary contractions,
\begin{equation}
\no{
{\color{sqred}
\contraction[1.05em]{}{\hat{q}}{ {}_{1} \qop{2} \cdots \no{} \qop{1}' \qop{2}'}{q}
}
{\color{sqred}
\contraction[0.75em]{\qop{1}}{\hat{q}}{ {}_{2}  \cdots \no{} \qop{1}' \!}{q}
}
{\color{sqred}
\contraction{\qop{1}\qop{2}}{\hat{q}}{\cdot\cdot \no{}  \!}{q}
}
\qop{1} \qop{2} \cdots 
}
\no{
\qop{1}' \qop{2}' \cdots 
},
\end{equation}
which reduces the total combinatorial factor by $1/3!$.

In the most general case, we consider a contraction involving $K$ operators characterized by operator matrices $\mathbf{N}_i= [\mathbf{n}_{i}^{+} \; \mathbf{n}_{i}^{-}]$ with $\mathbf{n}_{i}^{+} = (\np{1i},\ldots,\np{si})$ and $\mathbf{n}_{i}^{-} = (\nm{1i},\ldots,\nm{si})$, where $i = 1, \ldots, K$ labels operators and $s$ is the number of orbital subspaces.
The composite contraction connecting these operators is defined by the contraction matrices  $\mathbf{C}_{ij}= [\mathbf{c}_{ij}^{+} \; \mathbf{c}_{ij}^{-}]$, where $\mathbf{c}_{ij}^{+} = (c_{1,ij}^{+},\ldots,c_{s,ij}^{+})$ and $\mathbf{c}_{ij}^{-} = (c_{1,ij}^{-},\ldots,c_{s,ij}^{-})$ are the number of contractions of operator $i$ with the elementary contraction $j=1,\ldots,L$ in each subspace.
Then the combinatorial factor for a contraction takes the form
\begin{equation}
\frac{
\prod_{i = 1}^{L}
D(\mathbf{n}^+_i, \{ \mathbf{c}^+_{ij}\}_{j=1}^{L})
D(\mathbf{n}^-_i,\{ \mathbf{c}^-_{ij}\}_{j=1}^{L})
}{m_{\mathcal{C}_1}! m_{\mathcal{C}_2}!\cdots},
\end{equation}
where the function $D$ is defined in terms of the multinomial factor
\begin{equation}
D(\mathbf{n},\{\mathbf{c}_j\}_{j=1}^{L}) =
\prod_{k=1}^{s} 
\binom{n_k}{c_{k1},\ldots,c_{kL},n_k-\sum_j^L c_{kj}},
\end{equation}
while $m_{\mathcal{C}_j}$ is the number of times an elementary contraction $\mathcal{C}_j$ appears in the composite contraction being evaluated.

\bibliography{/Users/fevange/Documents/Research/Articles/Bibliography/papersappclean.bib,extra.bib}
\end{document}